\documentclass[10pt,twocolumn]{article}

\usepackage[T1]{fontenc}
\usepackage[utf8]{inputenc}
\usepackage[a4paper,margin=0.75in]{geometry}
\usepackage{amsmath}
\usepackage{amssymb}
\usepackage{graphicx}
\usepackage{bm}
\usepackage{xcolor}
\usepackage{authblk}
\usepackage{booktabs}
\usepackage{microtype}
\usepackage{caption}
\usepackage{cite}
\usepackage{flushend}
\usepackage[colorlinks=true,breaklinks=true]{hyperref}

\newcommand{\safeincludegraphics}[2][]{%
  \IfFileExists{#2}{\includegraphics[#1]{#2}}{%
    \fbox{\parbox{0.9\columnwidth}{\centering Missing figure file: \texttt{\detokenize{#2}}}}}}

\title{Small Clusters of $^4$He Atoms in Finite-Cutoff Effective Field Theory}

\author[1,*]{Betzalel Bazak}
\affil[1]{The Racah Institute of Physics, The Hebrew University, Jerusalem 9190401, Israel}
\affil[*]{\texttt{betzalel.bazak@mail.huji.ac.il}}

\date{}

\begin{document}

\twocolumn[
\begin{@twocolumnfalse}
\maketitle

\begin{abstract}
Small clusters of $^4$He atoms are benchmark systems for universal few-body physics near the unitary limit. We study these systems using a finite-cutoff effective field theory calibrated to low-energy observables from the realistic LM2M2 potential. The chosen two-body cutoff reproduces both the atom--atom scattering length and effective range, while a regulated three-body interaction is adjusted to the trimer and tetramer ground-state energies. We distinguish total binding energies from the one-atom separation energies $S_A^*=B_A^*-B_{A-1}$ of the shallow excited states. Ground-state energies through $A=8$ are reproduced at the few-percent level, but threshold-sensitive observables are less accurate. In particular, the calculated tetramer separation energy is $S_4^*=2.85$~mK, compared with $0.92$--$0.96$~mK in converged LM2M2 calculations, and the atom--trimer scattering length differs substantially from modern LM2M2 benchmarks. For $A\ge5$, the available excited-state benchmarks are sparse and strongly method dependent. Comparison with another soft-core interaction suggests that the discrepancies primarily reflect short-distance physics and higher-order operators omitted from the finite-cutoff Hamiltonian rather than numerical convergence of the stochastic variational calculation. Finite-cutoff EFT therefore provides an efficient description of ground-state systematics, whereas shallow excited states and atom--cluster scattering require a controlled higher-order and cutoff-dependence analysis.
\end{abstract}

\noindent\textbf{Keywords:}
Few-body systems; helium clusters; Efimov physics; effective field theory; universal physics

\vspace{1em}
\end{@twocolumnfalse}
]

\section{Introduction}
Clusters of $^4$He atoms have long served as benchmark systems for studying few-body quantum phenomena near the unitary limit. In particular, the observation of the excited $^4$He trimer state~\cite{Kun15} provided one of the first experimental confirmations of the Efimov effect~\cite{Efi70,NaiEnd17}. The weak binding of helium clusters makes them ideal testbeds for universal properties governed by a large two-body scattering length~\cite{BraHam06,GreGiaPer17}.

Recent advances in precision spectroscopy and diffraction techniques have enabled direct measurements of $^4$He dimer and trimer states~\cite{SchToe94,Kun15,Zel16,Voi14}, allowing quantitative comparison with high-accuracy theoretical predictions based on realistic helium pair potentials such as LM2M2~\cite{AziSla91} and PCKLJS~\cite{PrzCenKom10}. The close agreement between experiment and theory confirms the reliability of these interatomic potentials for low-energy helium systems.

Scattering processes involving bosonic helium clusters are strongly influenced by Efimov physics. For identical bosons, atom--dimer and dimer--dimer scattering observables are not determined by the atom--atom scattering length alone: they depend also on a three-body parameter and vary log-periodically as Efimov states cross the relevant thresholds~\cite{BraHam06,DelBosDD11}. Consequently, the constant ratios derived for dimers of two-component fermions, such as $a_{dd}\simeq0.6\,a_{aa}$, do not apply to $^4$He atoms. In the present work, atom--cluster scattering observables are therefore assessed by direct comparison with calculations for the bosonic LM2M2 Hamiltonian.

The Efimov effect, originally predicted for three identical bosons with resonant short-range interactions, manifests as an infinite series of geometrically spaced three-body bound states that persists even when the two-body subsystems are unbound~\cite{Efi70}. Beyond the three-body sector, universality extends to larger systems. For each Efimov trimer, two universal four-body states are expected, with properties determined by the trimer scale~\cite{PlaHamMei04,HamPla07,SteDInGre09}. Other observables, including the scattering lengths at which these states become unbound, also follow universal relations that have been confirmed experimentally in ultracold atomic gases near Feshbach resonances~\cite{PolDriHul09,DInSteGre09,Del10,FerKnoBer09}. Similar universal scaling has been identified in larger bosonic clusters~\cite{BluGre00,HanBlu06,Ste10,YamFedJen10,Ste11,GatKieViv12,GatKie14,YanBlu15,ZenHuaBer13,HigGre25}.

Effective field theories provide a systematic and model-independent framework for describing low-energy phenomena in terms of the relevant degrees of freedom, while encoding unresolved short-distance physics in contact interactions~\cite{Wei90}. They are particularly useful for universal systems, where observables depend only weakly on microscopic details~\cite{BraHam06}. In practical implementations, the EFT interaction is regularized with a momentum cutoff $\Lambda$, and renormalization is achieved when observables become insensitive to the cutoff within the expected truncation uncertainty.

At leading order (LO), the two-body potential is represented by a regulated contact interaction whose low-energy constant (LEC) is fitted to reproduce the two-body scattering length. At next-to-leading order (NLO), derivative operators are introduced to reproduce the effective range. Interestingly, at certain finite-cutoff values, the nominally LO interaction also reproduces the empirical effective range. In such cases, NLO-level accuracy may be achieved for selected low-energy observables without introducing additional derivative operators.

Working directly at such finite cutoffs provides a computationally efficient strategy, as proposed in Refs.~\cite{SchGirGne21,GatKie23,GatKieViv11,KieGarRom11,Del22,RecKieGir22,AvrBaz25}. Related interpretations of EFT calculations, which in practice are closely connected to this finite-cutoff strategy, have been discussed in Refs.~\cite{BeaFar22,ConSchGne25}.

Traditionally, calculations at several cutoff values are used to test renormalization-group invariance and validate the power counting. Such analyses have shown that the three-body force must be promoted to LO~\cite{KapSavWis98,BedHamKol99}, while a four-body force first enters at NLO~\cite{BazKirKon19}, in contrast to naive dimensional analysis. Once the power counting has been established, however, accurate and predictive results can be obtained using a single suitably chosen finite cutoff.

Varying the cutoff also provides a way to estimate residual cutoff dependence. Since this dependence is expected to be absorbed by higher-order contributions, its variation provides a lower bound on the theoretical uncertainty associated with omitted terms~\cite{Har20}. The present single-cutoff implementation does not provide this diagnostic. Because the selected finite cutoff also reproduces the effective range, we use $(r_{aa}/a_{aa})^2\simeq1\%$ as a nominal estimate of residual corrections for natural-size observables. As discussed below, this estimate does not translate directly into a relative uncertainty for small separation energies.

In this work, we apply finite-cutoff EFT to few-body helium systems. We tune the EFT interactions to reproduce low-energy observables obtained with the LM2M2 potential and solve the $A$-body Schr\"odinger equation up to $A=8$. This allows us to assess the accuracy and predictive power of the finite-cutoff EFT relative to realistic potentials and to predict atom--tetramer scattering parameters, extending the EFT description to larger helium systems.

The paper is organized as follows. In Sec.~\ref{sec:theory}, we describe the finite-cutoff EFT framework, the fitting procedure used to determine the low-energy constants, and the methods used to compute binding energies and scattering parameters. Section~\ref{sec:results} presents results for binding energies up to $A=8$, atom--cluster scattering parameters up to $A=5$, and structural properties of the cluster wave functions, together with comparisons to benchmark calculations based on realistic interatomic potentials. Section~\ref{sec:summary} summarizes our findings and outlines possible future extensions.

\section{Model and Methods}
\label{sec:theory}
We describe systems of $A$ helium atoms within finite-cutoff EFT. At LO, we employ regulated two- and three-body interactions,
\begin{equation}
  V = \sum_{i<j} V_2(r_{ij}) + \sum_{i<j<k} \sum_{\rm cyc} V_3(r_{ij},r_{jk})\,,
\end{equation}
where $r_{ij}=|\mathbf{r}_i-\mathbf{r}_j|$ is the relative distance between particles $i$ and $j$, and the three-body potential is summed over cyclic permutations of the triplet $\{ijk\}$.

\subsection{Two-body interaction}
The two-body potential is modeled as a Gaussian-regulated contact interaction,
\begin{equation}
V_2(r) = C_0\, \delta_{\Lambda_2}(r)\,,
\end{equation}
where $C_0$ is the two-body LEC and
\begin{equation}
\delta_{\Lambda}(r)=\exp(-\Lambda^2 r^2/4)\,.
\end{equation}
is an unnormalized Gaussian regulator with momentum cutoff $\Lambda_2$.

Rather than fitting directly to experimental data, we calibrate the EFT to low-energy observables obtained with the LM2M2 potential~\cite{AziSla91}. This strategy provides a controlled benchmark: the reference values are accurately known, while the comparison isolates the predictive content of the finite-cutoff EFT.

The parameters $C_0$ and $\Lambda_2$ are adjusted to reproduce the atom--atom scattering length $a_{aa}$ and effective range $r_{aa}$ extracted from LM2M2~\cite{AziSla91}. We adopt the mass convention $\hbar^2/m=12.11928$~K\AA$^2$ recommended in Ref.~\cite{RouCav12}. Many earlier calculations used the approximate value $\hbar^2/m=12.12$~K\AA$^2$, which leads to differences of order $0.5\%$ in the results~\cite{RouCav12}. The LM2M2 scattering observables used in this work are $a_{aa}=100.0$~\AA{} and $r_{aa}=7.33$~\AA{}~\cite{RouCav12}. The corresponding two-body LEC and cutoff are listed in Table~\ref{tab:LECs_Lambdas}.

\subsection{Three-body interaction}
At LO, a three-body counterterm is required to renormalize the trimer spectrum and avoid the Thomas collapse~\cite{Tho35}. We therefore include a regulated three-body contact potential,
\begin{equation}
V_3(r_{ij},r_{jk})
  = D_0\, \delta_{\Lambda_3}(r_{ij})\delta_{\Lambda_3}(r_{jk})\,,
\end{equation}
where $D_0$ and $\Lambda_3$ are the three-body LEC and cutoff. We determine these parameters through a joint calibration to the LM2M2 trimer and tetramer ground-state binding energies, $B_3=126.499$~mK~\cite{RouCav12,HiyKam12} and $B_4=559.22$~mK~\cite{HiyKam12}, respectively.

The resulting LECs and cutoff values are summarized in Table~\ref{tab:LECs_Lambdas}.

\begin{table}[t]
\centering
\caption{Low-energy constants and finite cutoffs employed in this work. The normalization of the Gaussian regulators is absorbed into the LECs.}
\label{tab:LECs_Lambdas}
\small
\setlength{\tabcolsep}{4pt}
\begin{tabular}{cccc}
\toprule
\multicolumn{2}{c}{LECs (mK)} & \multicolumn{2}{c}{Cutoffs (\AA$^{-1}$)} \\
\cmidrule(lr){1-2}\cmidrule(lr){3-4}
$C_0$ & $D_0$ & $\Lambda_2$ & $\Lambda_3$ \\
\midrule
$-1225.85$ & $653.28$ & $0.37658$ & $0.6$ \\
\bottomrule
\end{tabular}
\end{table}

Although different cutoffs are used in the two- and three-body sectors, both correspond to short-distance scales relative to the large atom--atom scattering length. Because the finite two-body cutoff is chosen to reproduce the effective range, a nominal estimate for residual corrections to natural-size total energies is $(r_{aa}/a_{aa})^2\simeq1\%$. This estimate cannot be assigned directly to a shallow separation energy $S_A^*=B_A^*-B_{A-1}$: a percent-level correction to either total binding energy may be comparable to, or larger than, their small difference. The parenthetical uncertainties quoted below quantify numerical convergence of the SVM calculation only and do not include EFT truncation or regulator uncertainty.

\subsection{Few-body calculations}
The few-body Schr\"odinger equation is solved in coordinate space using the stochastic variational method (SVM) with correlated Gaussian basis functions~\cite{SuzVar98,BazEliKol16}. This approach provides high numerical accuracy and efficiently captures both short-range correlations and the long-range behavior characteristic of weakly bound states.

Scattering observables are computed by placing the system in a harmonic trap of frequency $\omega$,
\begin{equation}
    V_{\rm HO}(\mathbf{r}) = \frac{m\omega^2}{2A} \sum_{i<j} (\mathbf{r}_i-\mathbf{r}_j)^2\,,
\end{equation}
and analyzing the discrete energy levels. The relation between the trapped spectrum and free-space scattering parameters follows Ref.~\cite{BusEngRza98}, which connects trap-induced energy shifts to the effective-range expansion of the scattering phase shift.

Consider scattering between two bound subclusters $B$ and $C$ in the trap. The relevant relative oscillator length is $b_{\rm rel}=\sqrt{\hbar/(\mu\omega)}$, where $\mu$ is the reduced mass of the two clusters. We choose $b_{\rm rel}$ to be much larger than all intrinsic length scales, so that the subclusters behave approximately as pointlike particles. The scattering channel is identified from the quantum numbers and energy ordering of the trapped states. For isolated levels, adiabatic continuation from weak to vanishing trap frequency associates a trapped level with the corresponding free-space scattering channel~\cite{BazEliKol16,SchBaz23}. The trapped solution is then matched to the analytic solution for two trapped particles with short-range interactions~\cite{BusEngRza98}. The $S$-wave phase shift $\delta$ at relative momentum $k$ is extracted from
\begin{equation}
    k \cot \delta = -\sqrt{\frac{4\mu \omega}{\hbar}}\,
    \frac{\Gamma\!\left[(3-2\epsilon)/4\right]}{\Gamma\!\left[(1-2\epsilon)/4\right]}\,,
    \label{Busch}
\end{equation}
where $\mu=m_B m_C/(m_B+m_C)$ is the reduced mass of the two clusters, $\Gamma(x)$ is the Gamma function, $k=\sqrt{2\mu\omega\epsilon/\hbar}$ is the relative momentum, and $\epsilon=(E_A-E_B-E_C)/(\hbar\omega)$ is the dimensionless energy of the trapped $A$-body system relative to the $B+C$ threshold. The trapped energies $E_A(\omega)$, $E_B(\omega)$, and $E_C(\omega)$ are computed with the SVM for several values of $\omega$. Substitution into Eq.~(\ref{Busch}) yields $k\cot\delta$ as a function of $k^2$.

The scattering length $a$ and effective range $r$ are then obtained by fitting the resulting values of $k\cot\delta$ to the effective-range expansion (ERE),
\begin{equation}
\label{eq:ere}
k \cot \delta = -\frac{1}{a} + \frac{1}{2}r k^2 + \mathcal{O}(k^4)\,.
\end{equation}
The trap-frequency range is chosen so that $k\cot\delta$ displays approximately linear behavior as a function of $k^2$ near threshold.

This formalism was used in Ref.~\cite{BazEliKol16} to determine atom--dimer scattering parameters of $^4$He atoms within LO EFT and was later extended to nuclear few-body systems, enabling few-nucleon scattering calculations at NLO~\cite{SchBaz23,BagSchBaz23,AvrBaz25}.

\begin{figure}[t]
  \centering
  \safeincludegraphics[width=\columnwidth]{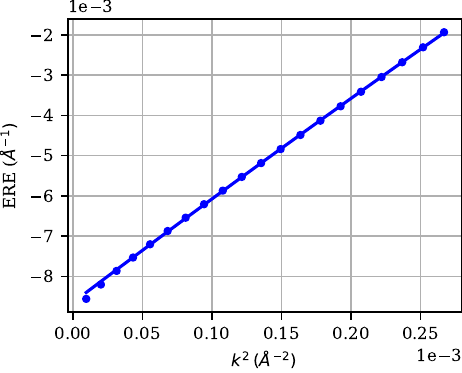}
  \caption{Effective-range expansion for $S$-wave atom--dimer scattering. Points are obtained from calculations in a weak harmonic trap using Eq.~(\ref{Busch}); solid curves show quadratic fits in $k^2$.}
  \label{fig:ad}
\end{figure}

As a representative example, Fig.~\ref{fig:ad} shows the ERE obtained for atom--dimer $S$-wave scattering using Eq.~(\ref{Busch}). The extracted points are well described by a quadratic polynomial in $k^2$, which also allows the stability of the threshold coefficients to be assessed.

\section{Results}
\label{sec:results}
Having fixed the EFT parameters, we now calculate the properties of helium clusters, including binding energies and atom--cluster scattering parameters.

\subsection{Two-atom system}
Because the two-body LECs are fixed solely by the atom--atom scattering length and effective range, the dimer binding energy $B_2$ provides the first prediction of the EFT. We find
\begin{equation}
B_2 = 1.3098~\text{mK}\,,
\end{equation}
in excellent agreement with the LM2M2 result $B_2=1.3094$~mK~\cite{RouCav12}. This confirms the internal consistency of the two-body sector.

\subsection{Three-atom system}
\begin{figure}[t]
  \centering
  \safeincludegraphics[width=\columnwidth]{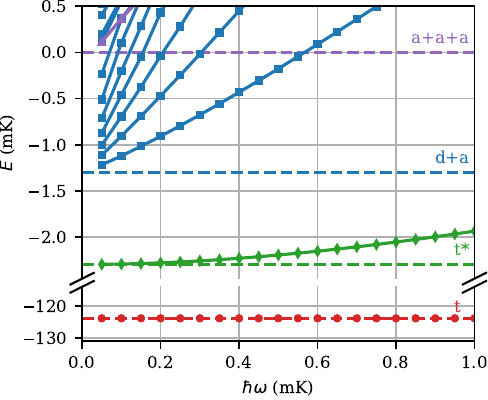}
  \caption{Energy spectrum of three $^4$He atoms in a harmonic trap as a function of the trap frequency. The nearly constant ground-state energy corresponds to the compact trimer, whereas the excited trimer shows stronger trap dependence due to its extended structure. Higher-lying states correspond to atom--dimer excitations and to the three-atom continuum. The free-space binding energies are indicated by dashed lines: $t$ for the deeply bound trimer, $t^*$ for the shallow trimer, $d+a$ for the atom--dimer threshold, and $a+a+a$ for the three-atom breakup threshold.}
  \label{He3_spec}
\end{figure}

The attractive helium interaction supports two bound trimer states. When three $^4$He atoms are confined in a weak harmonic trap and the trap frequency is varied, the ground-state energy remains nearly constant, reflecting its relatively compact spatial structure. By contrast, the excited trimer is weakly bound and spatially extended, and its energy shows a stronger dependence on the trap frequency, as shown in Fig.~\ref{He3_spec}. At higher excitation energies, the atom--dimer threshold is reached and atom--dimer-like trapped states appear. Their energies are related to atom--dimer scattering parameters through Eq.~(\ref{Busch}). The continuum of three unbound atoms appears at positive energies and displays characteristic avoided crossings with atom--dimer configurations.

The trimer ground-state energy has been extensively studied with a variety of few-body methods. Most calculations yield $B_3=126(1)$~mK~\cite{NieFedJen98,BluGre00,RouYak00,MotSanSof01,LazCar06,GuaKorNav06,SalYarLev06,HiyKam12,Del18,Del22,KorKol23}, although somewhat different results have also been reported~\cite{EsrLinGre99,GonRubMir99}.

We calibrate the three-body interaction jointly to the trimer and tetramer ground-state energies. These observables are correlated along the Tjon line~\cite{Tjo75}; therefore, although the three-body sector contains two adjustable parameters, $D_0$ and $\Lambda_3$, obtaining a consistent simultaneous description of both systems is nontrivial.

For the trimer ground state, we obtain
\begin{equation}
B_3 = 126.29(1)~\text{mK}\,,
\end{equation}
which differs by less than $0.2\%$ from the LM2M2 value $B_3=126.499$~mK~\cite{RouCav12,HiyKam12}.
For the excited trimer, we find
\begin{equation}
B_3^* = 2.3076(1)~\text{mK}\,,
\end{equation}
which differs by about $1.3\%$ from the LM2M2 result $B_3^*=2.2784$~mK~\cite{RouCav12,Del22}.

The atom--dimer scattering parameters are extracted from phase shifts obtained with Eq.~(\ref{Busch}), as illustrated in Fig.~\ref{fig:ad}. We obtain
\begin{equation}
a_{ad}=116.4(3)~\text{\AA}\,, \qquad r_{ad}=50.3(5)~\text{\AA}\,.
\end{equation}
The scattering length agrees well with the LM2M2 value $a_{ad}=115.39$~\AA{}~\cite{RouCav12,Del22}. For identical bosons, however, there is no symmetry-independent constant ratio $a_{ad}/a_{aa}$; this agreement should therefore be interpreted as a direct, system-specific benchmark. The extracted effective range is smaller than the LM2M2 result $r_{ad}=79.0$~\AA{} reported in Ref.~\cite{LazCar06}. This discrepancy may be related to the energy interval used in the fit, as Fig.~\ref{fig:ad} suggests a somewhat steeper slope at very low energies.

A comparison of available results for trimer binding energies and atom--dimer scattering parameters is given in Table~\ref{tbl:3He}. Most cited works use the LM2M2 potential directly. Ref.~\cite{PlaHamMei04} employs an EFT interaction, Ref.~\cite{GatKieViv11} uses a Gaussian soft-core interaction, and Ref.~\cite{Del22} solves a sequence of softened interactions and extrapolates to the LM2M2 limit.

\begin{table*}[t]
\centering
\caption{Properties of the three-atom system: ground- and excited-state trimer binding energies ($B_3$ and $B_3^*$), and atom--dimer scattering parameters ($a_{ad}$ and $r_{ad}$). Results above the horizontal line use LM2M2, either directly or through extrapolation to the LM2M2 limit; those below the line use effective interactions. Differences among literature values arise in part from different mass conventions; see text.}
\label{tbl:3He}
\small
\begin{tabular}{lcccc}
\toprule
Ref. & $B_3$ (mK) & $B_3^*$ (mK) & $a_{ad}$ (\AA) & $r_{ad}$ (\AA) \\
\midrule
\cite{BluGre00}    & 125.5(1)   & 2.187        & 125.9          & --             \\
\cite{MotSanSof01} & 125.9      & 2.28         & 131            & --             \\
\cite{BarKie01}    & 126.4      & 2.265        & --             & --             \\
\cite{Rou03}       & --         & --           & 115.4(1)       & --             \\
\cite{KolMotSan04} & --         & --           & 115.5(5)       & --             \\
\cite{LazCar06}    & 126.39     & 2.2680       & 115.56         & 79.0           \\
\cite{RouCav12}    & 126.499    & 2.27844      & 115.39         & --             \\
\cite{HiyKam12}    & 126.499    & 2.27787      & --             & --             \\
\cite{Del22}       & 126.50     & 2.2784       & 115.39         & --             \\
\midrule
\cite{PlaHamMei04} & 127        & --           & --             & --             \\
\cite{GatKieViv11} & 126.4      & 2.299        & --             & --             \\
This work          & 126.29(1)  & 2.3076(1)    & 116.4(3)       & 50.3(5)        \\
\bottomrule
\end{tabular}
\end{table*}

\subsection{Four-atom system}
The four-body sector provides an important benchmark for assessing the predictive power of the EFT\@.

The binding energies of $^4$He tetramers have been computed in numerous previous studies using realistic potentials~\cite{BluGre00,LazCar06,GuaKorNav06,HiyKam12}, effective interactions~\cite{GatKieViv11,PlaHamMei04,BazEliKol16,RecKieGir22}, and a softened-potential extrapolation to the LM2M2 limit~\cite{Del22}. A comparison is given in Table~\ref{tbl:4He}.

As discussed above, the LM2M2 tetramer ground-state energy, $B_4=559.22$~mK~\cite{HiyKam12}, is one of the two observables used in the joint three-body calibration. Our calculation gives
\begin{equation}
B_4 = 560.17(1)~\text{mK}\,,
\end{equation}
a deviation of less than $0.2\%$.
For the first excited tetramer state, we obtain
\begin{equation}
B_4^* = 129.14(1)~\text{mK}\,.
\end{equation}
Because this state lies immediately below the atom--trimer threshold, the physically relevant quantity is its one-atom separation energy,
\begin{equation}
S_4^* \equiv B_4^*-B_3 = 2.85(2)~\text{mK}\,.
\end{equation}
Converged LM2M2 calculations instead give $S_4^*=0.921$~mK from Ref.~\cite{HiyKam12} and approximately $0.96$~mK from Ref.~\cite{Del22}. Thus, although the total binding energy differs by only about $1\%$, the present Hamiltonian overbinds the shallow tetramer relative to the atom--trimer threshold by roughly a factor of three.

Only a few works have reported atom--trimer scattering parameters for the LM2M2 potential~\cite{BluGre00,LazCar06,Del22}. We find
\begin{equation}
a_{at}=69.7(1)~\text{\AA}\,, \qquad r_{at}=25.6(1)~\text{\AA}\,.
\end{equation}
The atom--trimer scattering length is tightly correlated with the position of the shallow tetramer pole. Consistent with the overestimated $S_4^*$, our $a_{at}$ differs substantially from the modern LM2M2 values $103.7$~\AA{}~\cite{LazCar06} and $108.8(5)$~\AA{}~\cite{Del22}, although it is closer to the older Monte Carlo hyperspherical estimate $56.1$~\AA{}~\cite{BluGre00}. The effective range is less discrepant. The quoted numerical uncertainties do not include EFT truncation or regulator effects and are far too small to account for these differences.

\begin{table*}[t]
\centering
\caption{Ground- and excited-state tetramer binding energies ($B_4$ and $B_4^*$) and atom--trimer scattering parameters ($a_{at}$ and $r_{at}$). Results above the horizontal line use LM2M2, either directly or through extrapolation to the LM2M2 limit; those below the line use effective interactions.}
\label{tbl:4He}
\small
\begin{tabular}{lcccc}
\toprule
Ref. & $B_4$ (mK) & $B_4^*$ (mK) & $a_{at}$ (\AA) & $r_{at}$ (\AA) \\
\midrule
\cite{BluGre00}    & 556.8(1)   & 132.7        & 56.1           & --      \\
\cite{LazCar06}    & 557.7      & 127.5        & 103.7          & 29.1    \\
\cite{GuaKorNav06} & 558(3)     & --           & --             & --      \\
\cite{HiyKam12}    & 559.22     & 127.42       & --             & --      \\
\cite{Del22}       & 559.3(1)   & 127.46(2)    & 108.8(5)       & 29.2(2) \\
\midrule
\cite{GatKieViv11} & 568.79     & 128.96       & --             & --      \\
\cite{PlaHamMei04} & 492        & 128          & --             & --      \\
\cite{RecKieGir22} & 560.206    & --           & --             & --      \\
This work          & 560.17(1)  & 129.14(1)    & 69.7(1)        & 25.6(1) \\
\bottomrule
\end{tabular}
\end{table*}

\subsection{Five-atom system}
The five-body system provides a further test of the EFT convergence pattern. Previous Monte Carlo calculations~\cite{BluGre00,GuaKorNav06,BazValBar20} yield $B_5\approx1300$~mK. Our EFT calculation gives
\begin{equation}
B_5 = 1292(1)~\text{mK}\,,
\end{equation}
in agreement with these benchmark results at approximately the $1\%$ level.

We also obtain a bound excited state at
\begin{equation}
B_5^* = 563.9(1)~\text{mK}\,,
\end{equation}
corresponding to the small separation energy $S_5^*=B_5^*-B_4=3.7(1)$~mK. This value is on the same few-mK scale as the $6.1$~mK separation inferred from the soft-core calculation of Ref.~\cite{GatKieViv11}, but it is much smaller than the $40.3$~mK value inferred from the LM2M2 Monte Carlo hyperspherical results of Ref.~\cite{BluGre00}. Because excited-state benchmarks for $A\geq5$ remain sparse and exhibit substantial method dependence, more accurate calculations are needed to clarify the near-threshold binding of this state. The available theoretical results and our predictions are summarized in Table~\ref{tbl:5He}.

Few calculations of atom--tetramer scattering parameters are available. We obtain
\begin{equation}
a_{a\tau}=56(1)~\text{\AA}\,, \qquad r_{a\tau}=18(1)~\text{\AA}\,.
\end{equation}
The scattering length is larger than the value $a_{a\tau}=33$~\AA{} reported in Ref.~\cite{BluGre00}.

When multiple cluster-scattering channels are present, such as atom--tetramer and dimer--trimer channels for $A=5$, trapped levels associated with different channels can undergo avoided crossings as the trap frequency is varied. In such regions, the states acquire mixed character and a coupled-channel generalization of the Busch formula would be required. Here, we extract the scattering parameters from spectral regions away from such avoided crossings, where a single-channel description is adequate. A coupled-channel treatment is left for future work.

\begin{table*}[t]
\centering
\caption{Ground- and excited-state pentamer binding energies ($B_5$ and $B_5^*$) and atom--tetramer scattering parameters ($a_{a\tau}$ and $r_{a\tau}$). Results above the horizontal line use LM2M2; those below the line use effective interactions.}
\label{tbl:5He}
\small
\begin{tabular}{lcccc}
\toprule
Ref. & $B_5$ (mK) & $B_5^*$ (mK) & $a_{a\tau}$ (\AA) & $r_{a\tau}$ (\AA) \\
\midrule
\cite{BluGre00}    & 1296(1)      & 597.1        & 33              & --  \\
\cite{GuaKorNav06} & 1310(5)      & --           & --              & --  \\
\cite{BazValBar20} & 1300(2)      & --           & --              & --  \\
\midrule
\cite{GatKieViv11} & 1326.6       & 574.9        & --              & --  \\
\cite{RecKieGir22} & 1292.7       & --           & --              & --  \\
This work          & 1292(1)      & 563.9(1)     & 56(1)           & 18(1) \\
\bottomrule
\end{tabular}
\end{table*}

\subsection{Larger clusters}
In the present calculation, the pattern of two bound states per cluster size, consisting of a deeply bound ground state and a shallow excited state close to the $(A-1)$-body threshold, persists for $A\ge6$. These systems have previously been studied with Monte Carlo methods~\cite{BluGre00,GuaKorNav06,BazValBar20}.

For the $A=6$ hexamer, we obtain
\begin{equation}
B_6 = 2268(1)~\text{mK}\,, \qquad B_6^* = 1301(1)~\text{mK}\,.
\end{equation}
For the $A=7$ heptamer, we find
\begin{equation}
B_7 = 3443(2)~\text{mK}\,, \qquad B_7^* = 2288(2)~\text{mK}\,,
\end{equation}
and for the $A=8$ octamer,
\begin{equation}
B_8 = 4781(3)~\text{mK}\,, \qquad B_8^* = 3477(3)~\text{mK}\,.
\end{equation}
Table~\ref{tbl:Energies} compares these EFT results for $A=6$--$8$ with previous calculations. The total ground-state binding energies remain within a few percent of the Monte Carlo benchmarks. However, the shallow-state separation energies are $S_6^*=9(2)$~mK, $S_7^*=20(2)$~mK, and $S_8^*=34(4)$~mK, and they do not exhibit comparably uniform agreement with the available LM2M2 estimates. Moreover, the available Monte Carlo estimates vary nonmonotonically with $A$, and extracting such small energy differences is numerically demanding because the excited states lie close to their respective $(A-1)+1$ thresholds. Close agreement for the much larger total binding energies is therefore insufficient to establish the accuracy of near-threshold states.

\begin{table*}[t]
\centering
\caption{Binding energies of small $^4$He clusters with $A=6$--$8$. Results above the horizontal line were obtained with the LM2M2 potential, whereas those below it use effective interactions. Energies are given in mK.}
\label{tbl:Energies}
\small
\begin{tabular}{lcccccc}
\toprule
Ref. & $B_6$ & $B_6^*$ & $B_7$ & $B_7^*$ & $B_8$ & $B_8^*$ \\
\midrule
\cite{BluGre00}    & 2309(3) & 1347    & 3565(4) & 2337    & 5020(4) & 3653    \\
\cite{GuaKorNav06} & 2308(5) & --      & 3552(6) & --      & 5030(8) & --      \\
\cite{BazValBar20} & 2315(2) & --      & 3571(2) & --      & 5041(2) & --      \\
\midrule
\cite{GatKieViv11} & 2338.9  & 1351.6  & --      & --      & --      & --      \\
\cite{RecKieGir22} & 2276    & --      & 3468    & --      & --      & --      \\
This work          & 2268(1) & 1301(1) & 3443(2) & 2288(2) & 4781(3) & 3477(3) \\
\bottomrule
\end{tabular}
\end{table*}

\subsection{Excited-state separation energies}
For a shallow $A$-body state, the physically relevant binding with respect to one-atom emission is
\begin{equation}
S_A^* = B_A^*-B_{A-1}\,.
\end{equation}
Table~\ref{tbl:separation} collects these separation energies. Each literature value is formed from energies obtained within the same calculation and mass convention. The $A=4$ benchmark is especially robust because independent Gaussian-expansion, Faddeev--Yakubovsky, and momentum-space calculations place $S_4^*$ near $1$~mK. The LM2M2 values for $A\ge5$ are taken mainly from the Monte Carlo hyperspherical estimates of Ref.~\cite{BluGre00} and should be regarded as less definitive, but they still expose the enhanced model dependence of small threshold differences.

\begin{table*}[t]
\centering
\caption{One-atom separation energies $S_A^*=B_A^*-B_{A-1}$ (mK) of the shallow excited states. Literature differences are constructed using the ground- and excited-state energies from the same reference. The LM2M2 values for $A\ge5$ are Monte Carlo hyperspherical estimates from Ref.~\cite{BluGre00}.}
\label{tbl:separation}
\small
\setlength{\tabcolsep}{5pt}
\begin{tabular}{cccc}
\toprule
$A$ & This work & LM2M2 & Soft-core model~\cite{GatKieViv11} \\
\midrule
3 & $0.9978(1)$ & $0.9690$~\cite{RouCav12,Del22} & -- \\
4 & $2.85(2)$   & $0.921$--$0.96$~\cite{HiyKam12,Del22} & $2.56$ \\
5 & $3.7(1)$    & $40.3$~\cite{BluGre00} & $6.11$ \\
6 & $9(2)$      & $51$~\cite{BluGre00}   & $25.0$ \\
7 & $20(2)$     & $28$~\cite{BluGre00}   & -- \\
8 & $34(4)$     & $88$~\cite{BluGre00}   & -- \\
\bottomrule
\end{tabular}
\end{table*}

The tetramer comparison demonstrates why a nominal percent-level description of total energies does not imply percent-level accuracy for a threshold observable. The present calculation and the independent calculation with a Gaussian soft-core interaction in Ref.~\cite{GatKieViv11} both give $S_4^*$ near $2.5$--$3$~mK, whereas realistic LM2M2 calculations give about $0.94$~mK. Since the SVM convergence uncertainty is orders of magnitude smaller than this discrepancy, the bound-state result is unlikely to be a numerical artifact. The similar behavior of the two soft-core Hamiltonians instead suggests sensitivity to short-distance physics not represented by the retained two- and three-body operators, including higher-order derivative and many-body interactions. The absence of the strong LM2M2 repulsive core is one manifestation of this difference, but within EFT its effects should ultimately be encoded in higher-order operators rather than by reproducing the core itself.

For $A\ge5$, the available benchmark separation energies are less precise and show sizable spread between realistic and soft-core calculations. We therefore cannot isolate a unique missing operator from the present data. Nevertheless, the results in Table~\ref{tbl:separation}, together with the discrepancies in atom--cluster scattering lengths, show that the current single-cutoff Hamiltonian is not quantitatively reliable for all shallow excited states.

\subsection{Structural properties}
As a further test of the EFT description beyond the energy spectrum, we compute the root-mean-square (rms) radius of each cluster,
\begin{equation}
r_{\rm rms} = \sqrt{\left\langle \frac{1}{A}\sum_{i=1}^{A}(\mathbf{r}_i - \mathbf{R}_{\rm cm})^2 \right\rangle}\,,
\end{equation}
where $\mathbf{R}_{\rm cm}$ is the center-of-mass coordinate. This quantity probes the spatial extent of the wave function and is sensitive to the balance between attractive two-body and repulsive three-body interactions.

The results are presented in Table~\ref{tbl:rms}, together with benchmark values from the Monte Carlo hyperspherical calculation of Ref.~\cite{BluGre00} and the Gaussian expansion method calculation of Ref.~\cite{HiyKam12}, both using LM2M2.

For the dimer, the rms radius of $35.4$~\AA{} is in excellent agreement with the Gaussian expansion method value of $35.47$~\AA{}~\cite{HiyKam12}, as expected for an extreme halo state whose size is governed primarily by the binding momentum rather than by the interaction range. For the trimer, our rms radius of $6.22$~\AA{} agrees well with the Monte Carlo hyperspherical value of $6.27(1)$~\AA{}~\cite{BluGre00} and the Gaussian expansion method value of $6.33$~\AA{}~\cite{HiyKam12}. For $A=4$, the EFT radius, $4.90$~\AA{}, is about $5\%$ smaller than the benchmark value, $5.16$~\AA{}. For $A\ge5$, the EFT and Monte Carlo radii agree within a few percent. The radii exceed the inverse cutoff scale $\Lambda_2^{-1}$, although the separation is only modest for $A\ge4$; the heavier clusters can therefore remain sensitive to short-distance details despite their spatially extended character.

\begin{table}[t]
\centering
\caption{Ground-state rms radii of $^4$He clusters. The last column gives the radius in units of the inverse two-body cutoff scale, $\Lambda_2^{-1}\approx2.66$~\AA.}
\label{tbl:rms}
\small
\setlength{\tabcolsep}{3pt}
\begin{tabular}{ccccc}
\toprule
$A$ & Ref.~\cite{BluGre00} (\AA) & Ref.~\cite{HiyKam12} (\AA) & EFT (\AA) & $r_{\rm rms}\Lambda_2$ \\
\midrule
2 & --      & 35.47 & 35.4 & 13.3 \\
3 & 6.27(1) & 6.33  & 6.22 & 2.34 \\
4 & 5.16(1) & 5.16  & 4.90 & 1.84 \\
5 & 4.90(1) & --    & 4.60 & 1.73 \\
6 & 4.85(1) & --    & 4.52 & 1.70 \\
7 & 4.86(1) & --    & 4.54 & 1.71 \\
8 & 4.92(1) & --    & 4.59 & 1.73 \\
\bottomrule
\end{tabular}
\end{table}

The dimensionless product $r_{\rm rms}\Lambda_2$ compares the cluster size with the inverse two-body cutoff scale. It is about 13 for the dimer but only $1.7$--$1.8$ for $A=4$--$8$. Thus, the dimer exhibits strong scale separation, whereas the heavier clusters retain only a modest separation between their rms size and the cutoff length. This observation is consistent with the increasing sensitivity of threshold observables to omitted short-distance operators.

\section{Conclusion}
\label{sec:summary}
We have applied a finite-cutoff EFT to few-body systems of $^4$He atoms. The two- and three-body interaction parameters were calibrated to reproduce the LM2M2 atom--atom scattering observables and to provide a simultaneous description of the trimer and tetramer ground-state energies. The resulting framework was then used to predict the properties of clusters containing up to eight atoms.

The EFT reproduces the large total ground-state binding energies with an accuracy ranging from sub-percent for the calibration systems to a few percent for the largest clusters. This success does not extend uniformly to shallow excited states. The appropriate one-atom separation energy of the excited tetramer is $2.85$~mK in the present calculation, compared with $0.92$--$0.96$~mK for LM2M2, and the associated atom--trimer scattering length correspondingly deviates substantially from modern LM2M2 benchmarks. For $A\ge5$, the available benchmark calculations are less uniform, but the separation energies and atom--cluster scattering lengths remain substantially more model dependent than the ground-state energies.

The computed rms radii show that all clusters from $A=2$ to $A=8$ are spatially extended relative to the inverse cutoff scale, with $r_{\rm rms}\Lambda_2\gtrsim1.7$. The scale separation is pronounced for the dimer but only modest for $A\ge4$, consistent with the enhanced short-distance sensitivity of the heavier near-threshold states.

The comparison also clarifies the scope of the nominal EFT uncertainty estimate. A correction of order $(r_{aa}/a_{aa})^2\simeq1\%$ to a total binding energy of order $10^2$--$10^3$~mK can be comparable to a separation energy of only a few mK. Relative uncertainties of shallow-state energies can therefore be of order unity even when the total energies appear accurate. The growth of ground-state deviations with $A$ may reflect the absence of a four-body force, which first enters at NLO in the adopted power counting~\cite{BazKirKon19}, while the threshold discrepancies may also receive enhanced contributions from derivative operators and regulator dependence. A multi-cutoff higher-order calculation is needed before attributing the discrepancy to a single operator.

These results show that finite-cutoff EFT is a practical framework for the ground-state systematics of weakly bound helium clusters, while also identifying an important limitation: observables governed by a pole extremely close to a breakup threshold are not controlled by the present single-cutoff, nominally LO Hamiltonian. Future work should include explicit higher-order operators and cutoff variation, and should test whether one additional few-body datum can stabilize the separation energies and scattering lengths. In addition, when multiple cluster-scattering channels lead to avoided crossings in the trapped spectrum, as in the atom--tetramer and dimer--trimer channels for $A=5$, a coupled-channel extension of the trapped scattering formalism will be required.

\section*{Acknowledgements}
This work was supported by the Israel Science Foundation (ISF), Grant No.~2441/24.

\section*{Declarations}

\noindent\textbf{Funding.}
This work was supported by the Israel Science Foundation (ISF), Grant No.~2441/24.

\noindent\textbf{Conflict of interest.}
The author declares that he has no conflict of interest.


\end{document}